\documentstyle[epsfig]{mn}
\newif\ifAMStwofonts

\title{Point source detection and extraction from simulated \\
 Planck TOD using optimal adaptive filters}

\author[Herranz et al.]{D. Herranz$^{1,2,}$\footnotemark, 
J. Gallegos$^{1}$, J.L. Sanz$^{1}$ 
 and E. Mart\'\i nez-Gonz\'alez$^{1}$ \\
 $^{1}$Instituto de F\'\i sica de Cantabria,  
             Fac. de Ciencias, Av. de los Castros s/n, 
             39005-Santander, Spain \\
 $^{2}$Departamento de F\'\i sica Moderna, 
             Universidad de Cantabria,       
             39005-Santander, Spain}

\date{\today}

\begin{document}

\maketitle

\begin{abstract}

Wavelet-related techniques have proven useful in the processing and analysis
of one and two dimensional data sets (spectra in the former case,
images in the latter). In this work we 
apply adaptive filters, introduced in a previous work (Sanz et al.
2001), to optimize the detection and extraction of point sources
from a one-dimensional array of time-ordered data such as the
one that will be produced by the future 30 GHz LFI28 channel 
of the ESA Planck mission.
At a $4\sigma$ detection level 224 sources over a
flux of 0.88 Jy are detected with a mean relative error
(in absolute value) of $21\%$ and a systematic bias
of $-7.7\%$. The position of the sources in the sky 
is determined with errors inferior to the size of
the pixel. The catalogue of detected sources 
is complete at fluxes $\geq$ 4.3 Jy. 
The number of spurious detections is less than a $10\%$ of the
true detections.
We compared the results with
the ones obtained by filtering with a Gaussian filter and a Mexican Hat
Wavelet of width equal to the scale of the sources.
The adaptive filter outperforms the other filters in all 
considered cases.
We conclude that optimal adaptive filters are well suited to
detect and extract sources with a given profile embedded
in a background of known statistical properties.
In the Planck case, they could
 be useful 
to obtain a real-time preliminary
catalogue of extragalactic sources, which
would have a great scientific interest, e. g. for follow-up
observations. 

\end{abstract}


\section{Introduction}
\footnotetext{e-mail: herranz@ifca.unican.es}

One of the most thrilling challenges in the study of the Cosmic
Microwave Background (CMB) is to deal with the problem of separating the
cosmological signal from the different foregrounds and
noises that appear in CMB experiments. 
This problem will be specially relevant in future high-resolution
experiments such as MAP (Bennett et al. 1996) and Planck (Mandolesi et al. 
1998, Puget et al. 1998). 
From the point of view of determining constraints on fundamental
cosmological parameters,
the different foregrounds (synchrotron emission,
galactic dust, free-free radiation,
thermal and kinetic Sunyaev-Zel'dovich (SZ) emission from galaxy
clusters and extragalactic point sources) are
considered as contaminants and therefore must be removed
together with the noise in order to extract the cosmological signal. Besides,
knowledge about each of these 'contaminants' has a great scientific relevance
on itself. Therefore, it is of exceptional importance 
to be provided with good techniques of denoising and foreground separation.

There are several methods already available to perform the separation
of the different components in CMB observations, such as the ones based
on Wiener filter (Tegmark and Efstathiou 1996, Bouchet and Gispert 1999)
and on Maximum Entropy Methods (MEM, Hobson et al. 1998, 1999; Stolyarov 
et al. 2001).
All these methods take advantage of the different
statistical properties
of the CMB and the foregrounds (e.g. different angular power
spectra)
as well as the knowledge of their distinctive frequency dependence.
When neither the spatial distribution of the foregrounds nor its 
frequency dependence are well known the performance of these 
separation methods
dramatically drops. Such is the case of point sources, whose spatial
distribution and abundance remains uncertain and whose frequency
dependence is not well known. Besides, they can show temporal and even
spectral variability. Extragalactic point sources should be
removed from the maps before any analysis should be performed.
Maximum Entropy Methods can produce a catalogue of point sources as
a residual (noise) from the separation process. However, only the
faintest sources are recovered, and the brightest ones can still
be observed in the residuals.
The Independent
Component Analysis technique (Baccigalupi et al 2000, Maino et al 2001) has also
been applied to this problem with promising results: it has the advantage that,
unlike the previous methods, it does not need any prior knowledge of the
components to be separated, but its  weakest point is actually the separation of
point sources.

Filtering techniques have been successful in both denoising and 
extracting the brightest point sources from CMB maps. Tegmark and
Oliveira-Costa 1998 presented a filtering scheme that minimizes
the variance of the map. However, their filter does not
produce a maximum gain (considered as the ratio between the $\sigma$-level
of the source after filtering and the $\sigma$-level of the source before
filtering) at the scale of the source. All the point sources in a CMB
map will have the same profile and size 
(the convolution of a $\delta$-Dirac source
with the antenna beam) and the filtering process should take advantage
of this fact using a  filter optimal for that particular profile and size
(scale). Wavelet techniques are specially suitable to deal with 
scales and spatial location at the same time. The application of the
Mexican Hat Wavelet (MHW) to realistic simulations was presented in Cay\'on
et al. (2000) and extended in Vielva et al. (2000a).
The principal advantage of this method is that no specific
assumptions about the statistical properties of the point source
population or the underlying emission from the CMB and other foregrounds
is required. The MHW and MEM techniques have successfully been combined to
extract both faint and bright point sources at the same time (Vielva et al.
2000b).
 
It is desirable to exploit to the maximum the mentioned scale
distinctiveness of point sources. In Sanz et al. 2001 (S01,
hereafter), a method was presented to, for a given source brightness
profile and given the statistical properties of the background (CMB
plus foregrounds and noise), derive the analytic shape of the 
optimal filter for the particular considered case. That kind of
filter was called 'optimal pseudo-filter'. 
S01 defines as an optimal pseudo-filter one that a) is unbiased, b) gives
a maximum at the position and scale of the source and c) gives
the minimum variance of pseudo-filter coefficients 
(i.e. is an efficient estimator of the amplitude of the sources)
under assumptions a) and
b). They considered one, two and three-dimensional cases and showed
that the MHW is optimal (in the mentioned sense) or nearly optimal
for a two-dimensional Gaussian profile and reasonable 
foreground conditions.
After carefully thinking about the terminology used in
that previous work, we have decided
to substitute the rather confusing 'pseudo-filter' term by
{\it 'adaptive filter'}. This is so because a filter that satisfies conditions a) to c)
{\it adapts} itself to the characteristics of the signal and 
the noise. In the following we will use 'adaptive filter' in the same sense 
as 'pseudo-filter' in S01.

Although two-dimensional maps are the most useful and extended form
to show CMB data, they are only available after an exhaustive
process of analysis and reduction of the raw data from CMB
experiments. These experiments scan different patches of the sky
in a sequence producing a unidimensional set of time ordered data (TOD).
TODs suffer from  lower signal-to-noise ratios than final two-dimensional
maps, but on the other hand are less likely to show artifacts coming
from the data reduction, such as pixel-to-pixel noise correlations.
Moreover, TODs can be analyzed in (almost) real time during the observations
in order to produce early (preliminary) catalogs of sources. 

In this paper we apply the techniques developed in S01 to 
realistic simulations of the TOD coming 
from one of the 30GHz Planck LFI's  
channels. In 
section~\ref{formalism} we summarize some of the conclusions of S01
and present the semi-analytic adaptive filter that should be used in
a realistic case. Section~\ref{data} describes the simulations
used in this work. 
In section~\ref{analysis} we describe the analysis of the simulated
data.
In section~\ref{results} we describe the performance of 
the optimal adaptive filter, comparing it with other filtering 
schemes such as Gaussian filter 
and MHW. Finally, in section~\ref{conclusions}
we discuss our conclusions and give an outline of future work in this 
field.

\section{One-dimensional adaptive filter} \label{formalism}

The general formalism of adaptive filters in an n-dimensional image was
presented in S01. TODs can be considered as a particular case of a 
one-dimensional
image (a spectrum is another interesting case). The image data values can be
expressed as
\begin{equation}
y(t)=s(t)+n(t),
\end{equation}
\noindent
where $t$ is the time, $s(t)$ represents a symmetric source
and $n(t)$ is a homogeneous and isotropic background with mean value 
$<n(t)> \, =0$ and characterized by the power spectrum $P(q)$ ($q$ is the
absolute value of the  
'wave vector' associated to $t$). If $A=s(0)$ is the amplitude of
the source we can introduce the profile $\tau(t)$ as $s(t)=A\tau(t)$.
Let us introduce a centrally symmetric
adaptive filter $\Psi(t;b,R)$
\begin{equation}
\Psi(t,b,R)= \frac{1}{R}\psi(\frac{\vert t-b \vert}{R})
\end{equation}
\noindent
where $b$ defines a translation whereas $R$ defines a scaling. The
adaptive filtered field is defined as:
\begin{equation}\label{filtered}
w(R,b)=\int dt \, y(t) \Psi(t;b,R) \equiv \int dq \, e^{-iqt}y(q)\psi(Rq)
\end{equation} 
\noindent
In the last equivalence we have expressed the convolution as a product 
in Fourier space.

Following S01, it is possible to derive the analytic form 
of the {\it optimal} adaptive filter
for a
given source profile, once we have defined what we mean by {\it 'optimal'}.
Let us define an {\it optimal} adaptive filter as one that satisfies
the following conditions:
\begin{enumerate}
\item There exists a scale $R_{o}$ such that at the point source position 
($b=0$) $<w(R,0)>$ has a
maximum at that scale. 
\item $<w(R_{o},0)>=s(0)$, i.e. $w(R_{o},0)$ is an unbiased estimator of
the amplitude of the source.
\item The variance of $w(R,b)$ has a minimum at the scale $R_{o}$, i.e. we have
an efficient estimator.
\end{enumerate}

Given these three conditions, the solution (adaptive filter) is
found to be:
\begin{equation}\label{eqfilter}
\tilde{\psi}(q) \equiv \psi(R_{o}q) = \frac{\tau(q)}{2P(q)\Delta}[b+c-
(a+b)\frac{d\ln\tau}{d\ln q}]
\end{equation}
\begin{eqnarray}\label{integrals}
a \equiv \int dq \frac{\tau^2}{P}, \ \
b \equiv \int dq \frac{\tau}{P}\frac{d \tau}{d\ln q}, \nonumber\\
c \equiv \int dq \frac{1}{P}[\frac{d \tau}{d\ln q}]^2, \ \ 
\Delta = ac-b^2 
\end{eqnarray}
\noindent
The limits of the integrals go from $0$ to $\infty$. 

In S01 analytic expressions for Gaussian sources and backgrounds of
the type $P(q) \propto q^{-\gamma}$ were derived. In a more realistic
case, the background can not be modelled in such a simple way, and 
integrals in (\ref{integrals}) should be numerically estimated. When dealing
with real data (or realistic simulations such as those used 
in this work) we must perform the following steps: first, determine the
power spectrum of the background directly from the image. Second,
evaluate integrals  (\ref{integrals}). Third, build the adaptive filter
(\ref{eqfilter}) and make the convolution (\ref{filtered}). Finally,
we can proceed to detect the sources, for example looking for
peaks above a certain $\sigma$-level in the coefficient (filtered) image.

  When determining the power spectrum of an image we
obtain the power spectrum of both the background and the sources
together. In the following we consider that the contribution of the
point sources to the total power spectrum is negligible. This is a
reasonable 
assumption in a realistic case, specially at medium and high
wavelengths where the emission of IR and radio sources is weak.
Another problem related to power spectrum determination is the
variance of the power spectrum estimator. The variance of
one-dimensional power spectrum estimators tends to be larger than
that of two-dimensional cases because
the number of sample points is usually  smaller. A possible 
solution to this problem is to estimate low resolution power spectra.
However, 
a large amount of information is lost at scales that are crucial
for the determination of the adaptive filters. For this work we
chose to average the estimated power spectra of adjacent rings of the
TOD (corresponding to regions of the sky separated by only a 
few arcminutes and therefore possessing very similar underlying
power spectra).

\begin{figure}
\epsfxsize=84mm
\epsffile{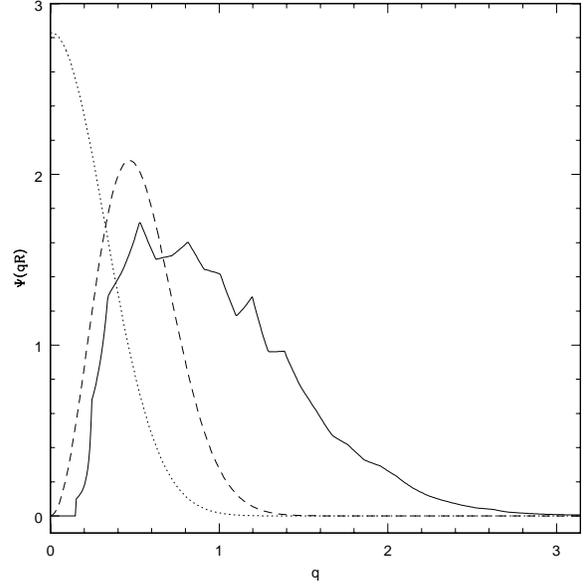}
\caption{Several filters in Fourier space. 
The optimal adaptive filter
particularised for an individual ring of our simulations 
is represented by a solid line. The dashed line 
shows a Mexican Hat Wavelet of width equal to
the width of the (Gaussian) source. The dotted line
represents a Gaussian of width equal to the width 
of the source. All filters are normalized to give
the true amplitude of the source after convolution.}
\label{fig1}
\end{figure}

   A typical profile of the optimal adaptive filter (in
Fourier space)
for a section of the simulated TOD used in this paper
 is shown in figure~\ref{fig1}. The profile (solid line) 
is irregular due 
to the roughness of the estimated power spectrum. 
These irregularities reflect the particularities
of the data and define the scales where the adaptive filter
is more or less efficient. For comparison, a Mexican Hat
Wavelet  (dashed line) and a Gaussian (dotted line),
both of them with a width equal to the width of
the source, are also shown.

\section{Data}\label{data}

For this work we chose the 30 GHz LFI28 channel of Planck 
because of the relatively small
size of its resulting TOD. Higher frequencies, in spite 
of being expected to show more contribution from point
sources due to their higher resolution and the contribution of
IR sources at $\nu \geq 300$ GHz, lead 
to huge TOD sets and require many hours 
of computation to be
simulated and analysed
on an average computer  and will be analyzed in a future work.

Our data simulates a 6 months run of the 30 GHz LFI28 
channel of Planck, covering  the whole sky except
for two circles of $1 \degr 82$ around the ecliptic poles.
The sky simulation consist of the addition (in flux) of a CMB simulation and
three templates, each for the Galactic synchrotron radiation, Galactic dust and
thermal SZ.
These templates have been generated using 
proper models that are
freely distributed among the Planck 
collaboration\footnote{http://planck.MPA-Garching.MPG.DE/SimData/}.
 This sky is then combined with a map of point sources, generated
from a catalog performed following the model of Toffolatti et al. 1998
(that is not public for the Planck collaboration)
 and then 'observed' by the Planck Pipeline
Simulator for one of the LFI 30 GHz radiometers; resulting in a TOD for the
observed sky (CMB, foregrounds and point sources) plus instrumental noise.
The data are ordered in circular rings centered on points
situated on
 the Ecliptic. 
The angle between the pointing axis and the rotation
axis for the LFI28 instrument is $88 \degr 18$.
The maximum separation between two consecutive rings
is $2.5 \arcmin$ (at the intersection with the Ecliptic plane).
Each ring results from the average of 60 revolutions
of the detector around the rotation axis, corresponding
to one hour of integration time, and contains 1950
measures of antenna temperature. There are a total of 4383 of such 
rings,
leading to 8546850 measures of temperature. The antenna has a 
FWHM of $33 \arcmin$ and its response slightly differs from a circular
Gaussian one.

\begin{figure*}
\epsfxsize=170mm
\epsffile{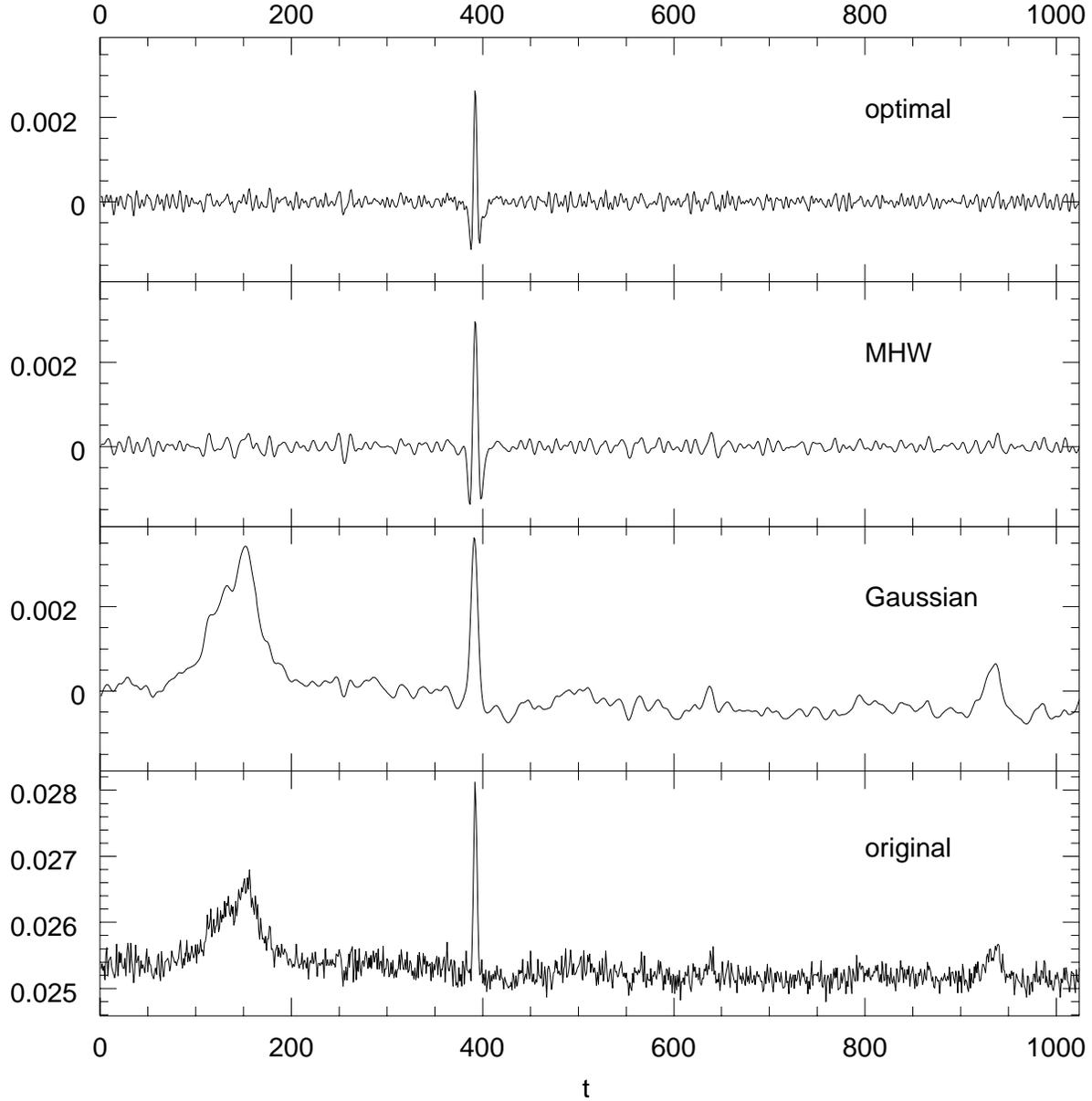}
\caption{A 1024-pixel section of one
of the rings before and after filtering. The lowest
panel shows the section before any filtering. The second 
panel shows the section after filtering with a Gaussian
of FWHM $33\arcmin$ (equal to the one of the source).
The third panel shows the section after filtering
with a Mexican Hat Wavelet of width equal to the
width of the source. The top panel shows the
section after filtering with the optimal adaptive filter.} 
\label{fig2}
\end{figure*}

The simulation contains CMB emission, different 
extended foregrounds 
(Galactic synchrotron, dust and free-free, thermal and kinetic SZ 
emission from clusters, etc),
point sources and instrumental noise. 
Both white noise and $1/f$ noise are present. The knee frequency
$f_{\nu}$ is set to be $<20$ mHz (less than
the frequency of rotation). In the lower
panel of figure~\ref{fig2} we show a segment
of one of the rings of the simulation. There is
a bright point source near pixel 400 (in fact,
it is the brightest source in the simulation).
Apart from this extraordinarily bright source,
the features that dominate the image are 
Galactic emission (the large peaks around
pixels 150 and 940) and noise.

\section{Data analysis}\label{analysis}

The complete set of simulated TOD
was filtered using the optimal adaptive filter described in
section~\ref{data}. Each individual ring was filtered
separately. The power spectrum that appears in equations
(\ref{eqfilter}) and (\ref{integrals}) was obtained by
averaging the estimated power spectra of twenty-one 
consecutive rings (the ring that is being filtered,
the ten previous rings and the ten subsequent
rings). 

In order to detect the sources from the
filtered image one can set a certain threshold
over the dispersion of the filtered ring
and look for connected regions (peaks) above that threshold.
That would be the most direct detection method if
the different rings were independent of their neighbours.
This is not the case, since adjacent rings scan very  
close regions and each source is expected to appear
in more than one ring. The use of information coming
from neighbouring rings allows to increase the
effective signal to noise ratio of the detections and
to discard spurious 'sources' due to noise 
fluctuations. The most straightforward way to
detect sources is then to perform a kind of two-dimensional
thresholding, looking for connected pixels at equal 
latitude that appear {\it in several adjacent rings}.  
In a first approximation, the
position of the source will be the position of the
maximum of that region of connected pixels. We will show
that this approximation is good enough for the purpose 
of locating the sources with an error comparable to
$1/3$ of the antenna FWHM.

   Two different regimes of 'noise' have to be removed
in order to optimize the detection of the sources. The
large scale features due to Galaxy foregrounds as well
as CMB fluctuations are strongly correlated between a
given ring of data and its neighbours. On the other hand,
the small scale noise, dominated by white instrumental
noise, is expected to be nearly independent from one
ring to another. 
This suggests a further step in the
idea of combining information from nearby rings to 
increase the signal to noise ratio of the sources. By averaging
each ring data (before filtering) we can construct a 'synthetic
TOD' in which the large scale fluctuations are almost the
same than in the original TOD but the noise at the scale
of the pixel has been greatly diminished. The $1/f$ will still
be present in the averaged TOD.
The point sources of the original TOD are replaced in the 
averaged TOD by Gaussians of amplitude
\begin{equation}\label{average}
A_{a}=\frac{1}{N_{a}} \sum_{i=1}^{N_{a}} A e^{-\frac{\mid \vec{x_{o}} -
  \vec{x_{i}} \mid^{2}}{2\sigma_{b}^{2}}}
\end{equation}  
\noindent where $A$ is the true amplitude of the source, $N_{a}$
is the number of rings that are averaged,
$\sigma_{b}$
is the beam width and the
distance $\mid \vec{x_{o}} -
  \vec{x_{i}} \mid^{2}$ is the geodesic (spherical) angular distance 
between the pixel corresponding to the position of the source and
the pixel that is being added to calculate the average.
This is true when the beam is a perfect Gaussian because in that case
the average is a weighted sum of Gaussians of equal width, that is, 
a new Gaussian of the same width and amplitude given by eq. (\ref{average}).
Therefore, we can filter the rather denoised, 'synthetic TOD', instead
of the original TOD and correct the amplitude of the sources
that we will detect using eq. (\ref{average}) in order to 
recover an estimate $\hat{A}$ of the true amplitude $A$.

Although the Planck Pipeline Simulator uses a realistic beam, we
assumed for this work that the beam is a perfect Gaussian. 
In section~\ref{results} it will be shown that this approximation
works reasonably well. However,
future work should take into account real beams; the 30 GHz detector
is expected to be slightly elliptic. This asymmetry will introduce
complicate effects in the map making and analysis. In the case of
TOD the effect of ellipticity, when projected over the
scan trajectory, is a change in the `effective width' of the
beam. Therefore, the scale-adaptive filter should be calculated
for such `effective width'. More complicated asymmetries 
will require more careful treatment. Future work will deal
with this issue.

\section{Results}\label{results}

In order to determine the goodness of the averaging of close
rings we performed two different filterings of the TOD. In the
first case, the raw data of the TOD were filtered ring by 
ring as described in the section~\ref{analysis}. In
the second case, a 'synthetic TOD' in which each individual ring
was the average of the original ring (at that position) with 
the twelve neighbours
was constructed. 
The number of rings averaged is such that at the maximum ring
spatial separation region on the sky (the ecliptic Equator)
the separation between the two most separated rings is approximately
equal to the FWHM of the beam. 
After filtering, the detection and extraction was 
performed looking for sets of 5 or more connected pixels (in 
the 2-dimensional sense). A lesser number of  connected
pixels required would lead to a much higher number of spurious detections.
To check if the detections correspond
to real sources we
compare with the catalogue of the 1000
brightest sources present in the simulations. This
catalogue is complete and its
flux limit  is clearly lower than
 the expected flux limit of the
detected sources.

\begin{figure}
\epsfxsize=84mm
\epsffile{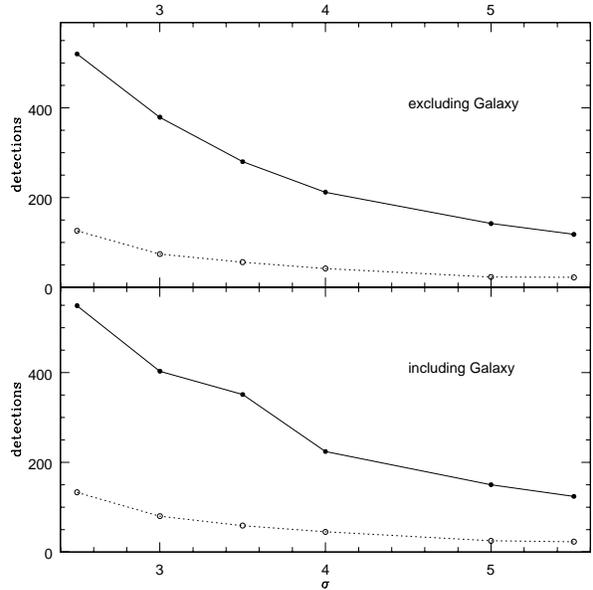}
\caption{30 GHz detections with the optimal adaptive filter
as a function of the $\sigma$ detection threshold.
The open circles and dashed lines represent detections over
the filtered TOD. The filled circles and solid line represent
detections over a filtered synthetic TOD in which each ring
is the result of the average of 9 rings of the original TOD. 
The synthetic TOD has the same number of rings than the 
original one. Two cases have been represented: simple 
detection over a certain threshold (lower panel) and  simple
detection excluding a $5\deg$ band around the Galactic plane
(top panel).}
\label{fig3}
\end{figure}

In the lower panel of figure~\ref{fig3} the number of detections (defined as 
the number of peaks encountered above a certain threshold
that correspond to a real source in the reference catalog)
above several $\sigma$ thresholds is shown 
for the case where the data have been averaged (filled circles and
solid line) and the case where they have not (open circles and dashed
line).
The remaining quantities of interest
(such as estimation of the amplitudes, fluxes, etc.)
concerning the detection/extraction of sources over the
averaged and then filtered TOD are shown in table~\ref{tb1} and will be
discussed later.  As expected, the number of sources detected when
analysing the synthetic TOD
is 4 to 5 times higher to
the number of sources detected over the original, filtered TOD for
every $\sigma$ level. 

\begin{table*}
 \begin{center}
 \begin{minipage}{122mm}	
  \caption{\label{tb1} 
Detections at 30GHz with the Optimal Adaptive filter,
 compared 
to a Mexican Hat Wavelet (MHW) and a Gaussian window 
with FWHM of $33 \arcmin$. 
Col. (1): $\sigma$ detection level. Col. (2): 
Number of sources found. Col. (3): Number of
spurious detections. Col. (4): Mean position offset. Col. (5): 
Mean relative absolute error of 
the amplitude (defined as $r.a.e. =
\mid A_{0} - A_{e}\mid/A_{0}$, where $A_{e}$ is the estimated amplitude).
Col. (6):
Mean bias in the amplitude. Col. (7): 
Minimum
reached flux. Col. (8): Flux over which the catalog of detections is complete. 
 }

 \begin{tabular}{ c c c c c c c c }
\hline
  $\sigma$ & 
  detected & 
  spurious & 
  mean offset   & 
  m.r.a.e.  & 
  $<bias>$ &
  min. flux  & 
  compl. flux \\
	&
  sources &
  sources &
($\arcmin$) &
($\%$) &
($\%$) &
 (Jy) &
 (Jy) \\

\hline
\\

 \multicolumn{7}{c}{Optimal Adaptive filter} \\ 

\\

2.5  &    549    &  1368  &   14.08   &  24.22 & 3.15  &  0.542 &  4.337  \\ 
3.0  &    403   &   296  &    13.56   &  24.32 & 1.31   & 0.644  & 4.337 \\ 
3.5   &   351    &  54    &   12.73   &  21.75 & -4.91   &  0.760   & 4.337\\ 
4.0   &   224    &  22     &  12.49   &  20.98 & -7.69  &  0.885  &  4.337\\ 
5.0   &   150    &  9      &  12.20   & 19.82&	-14.85   & 1.065   & 4.337\\ 
5.5   &   124     &  5      &  12.03   & 20.38 & -15.54  &  1.197   & 4.337\\ 

\\

 \multicolumn{7}{c}{Gaussian filter} \\ 

\\

2.5  &    11 &      11 &      18.88 & 538.4 & 	 537.0 &  6.511  &  17.070 \\
3.0  &    7  &      6  &      17.84 & 746.4 & 	 744.2 &  8.787  &  10.070 \\
3.5  &    7  &      13 &      17.50 & 898.1 &  	898.1 &  7.908  &  18.866 \\
4.0  &    4  &      11 &      17.24 & 1281.7 & 	1281.7 &  10.340 &  18.866 \\
5.0  &    4  &      12 &      13.09 & 1985.3 &  1985.3 &  13.926 &  18.866 \\
5.5  &    4  &      15 &      13.09 & 1985.3 &  1985.3 &  13.926 &  18.866 \\

\\

 \multicolumn{7}{c}{Mexican Hat Wavelet} \\ 

\\

2.5  &    473  &    413  &    12.83 & 21.46 &	4.27 &   0.531 &   4.337\\
3.0  &    361  &    130  &    12.66 & 20.76 &	2.52 &   0.693 &   4.337\\
3.5  &    270  &    63   &    12.67 & 20.46 & 	0.45  &   0.761 &   4.337\\
4.0  &    196  &    59   &    12.40 & 18.47 &  -4.22  &   0.844 &   4.337\\
5.0  &    139  &    49   &    12.50 & 19.47 &  -4.96  &   0.957 &   4.337\\
5.5  &    118   &   44   &    12.41 & 19.83 &  -5.53  &   1.099 &   4.337\\

\hline

 \end{tabular}
\end{minipage}
\end{center}
 \end{table*}

\begin{figure}
\epsfxsize=84mm
\epsffile{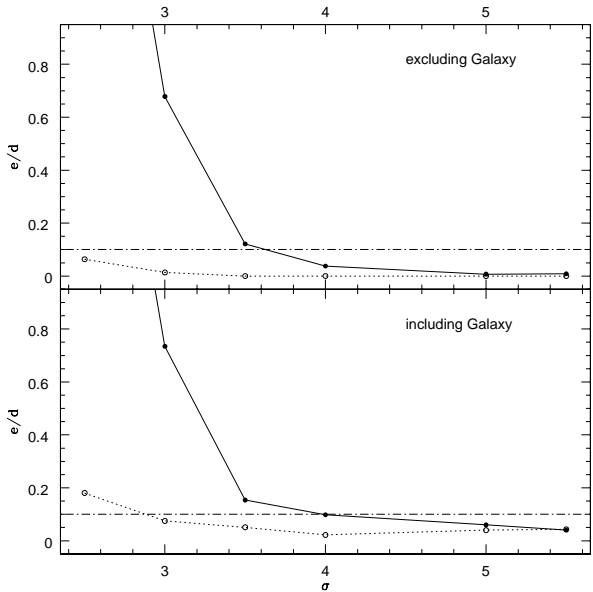}
\caption{30 GHz ratio between the number of spurious sources
and the number of true detections with the optimal adaptive filter
as a function of the $\sigma$ detection threshold.
The open circles and dashed lines represent detections over
the filtered TOD. The filled circles and solid line represent
detections over a filtered synthetic TOD in which each ring
is the result of the average of 9 rings of the original TOD. 
The synthetic TOD has the same number of rings than the 
original one. Two cases have been represented: simple 
detection over a certain threshold (lower panel) and simple
detection excluding a $5\deg$ band around the Galactic plane
(top panel).}
\label{fig4}
\end{figure}

   Apart from having the greatest number of detections, it
is also of great importance to reduce as much as possible the
number of spurious detections. In the lower panel of figure~\ref{fig4}
(labeled as 'simple detection')
the ratio between spurious and 'true' detections is represented 
for the case where the data have been averaged (filled circles and
solid line) and the case where they have not (open circles and dashed
line). The ratio is lower for the case in which the data have not been
averaged. It is evident that a kind of compromise has to be reached
between gain and reliability. If we arbitrarily set a maximum proportion
of spurious sources versus true ones, say a $10\%$ (represented in
fig.~\ref{fig4} with an horizontal  dashed line), we can determine 
the minimum $\sigma$ level that satisfies this condition. In this example, 
for the case of non-averaged rings, we can reach the $3\sigma$ and find
80 sources (to a minimum flux of 4.33 Jy) 
with a $7.5\%$ of spurious detections. For the case of averaged
rings, we must go to the $4\sigma$ level, where we
find 224 sources (to a minimum flux of 0.885 Jy) with a
$9.8\%$ of spurious detections.  We conclude that the averaging
of neighbouring rings is a valid strategy to reduce pixel-scale noise.
In the following, all the results will refer to filtering
of averaged rings.

 The number of detections and  spurious sources found with
the optimal adaptive filter applied to a synthetic TOD  is
shown in table~\ref{tb1}. The determination of the position of the
source, the mean relative absolute error in the determination of the amplitude
(defined as $m.r.a.e. =
\mid A_{0} - A_{e}\mid/A_{0}$, where $A_{e}$ is the estimated amplitude and
$A_{0}$ is the real amplitude), 
the mean bias (defined in the same way as the m.r.a.e, but without the
absolute value),
the minimum flux reached and
the completeness flux are also included in table~\ref{tb1}. 
In each case the mean error in the 
position of the sources is comparable with the size of
the 'pixel' of the TOD, $11\arcmin$. The determination of 
the amplitude using eq. (\ref{average}) has relative errors
ranging from $24.32 \%$ at $3\sigma$ threshold to $19.82\%$ at
$5.0\sigma$. The error decreases as the detection threshold increases.
This indicates that the estimation of the amplitude of weak sources is
less accurate than the estimation of the amplitude of bright ones. 
Also, the determination of the amplitude 
is biased to 
higher values at low $\sigma$ levels and to lower values
(negative bias) at high $\sigma$ levels.
 The positive bias
for weak sources arises due to the peak finding algorithm:
it finds preferently the maxima in pixels where the 
noise contribution is positive. 

Let us consider for a while the possible causes for the mentioned
bias and how to deal with it. 
Scale-adaptive filter is designed to be an unbiased
estimator of the amplitude of the sources. Thus, \emph{the
method is unbiased}. However, in practise it is found a
small bias. The origin of this bias can be found in two different
kind of effects. 
On the one hand, the bias introduced by the detection method:
this bias was described in the last paragraph and is related
with the well-known `detection bias'. On the other hand, bias
may arise due to the non-ideality of
the data. In first place, the profiles considered in the design of
the filters are continuous whereas real data is pixelised. Therefore,
the correlation between the (continuous) filter and the (pixelised)
source profile is not perfect, this resulting in a non-ideal
performance of the filter. In second place, the limits in 
the integrals 
(\ref{integrals}) are from $0$ to $\infty$, whereas in real
data the frequencies are limited by the sizes of the
data set and the pixel. Therefore,
 integrals (\ref{integrals}) can only be approximately calculated.
This leads to an inaccuracy in the determination of the shape of the
filter and, more important for the bias, the \emph{normalization} 
(the $\Delta$ in the mentioned integrals). Another source of bias
appears when the assumption about the profile of the sources is
wrong. For example, we have assumed that the detector beam is Gaussian,
while in fact it is not; this could explain the bias in our
results. 
In SO1 it was found using simple simulations 
that the bias due to the non-ideality of
the data was negative. Similarly,
the negative bias that appear here is interpreted 
in the way described above.
In practise, it is difficult to distinguish from
the data the different
contributions to the bias among the ones above mentioned. 
A possibility to overcome the bias is to calibrate it using
a great number of simulations with 
the same background and artificial sources of known amplitude.
Once removed the `systematic' bias, the remaining
error will be statistical.
This kind of study will be carried on in a future work.

\begin{figure*}
\includegraphics[angle=270, width=17cm]{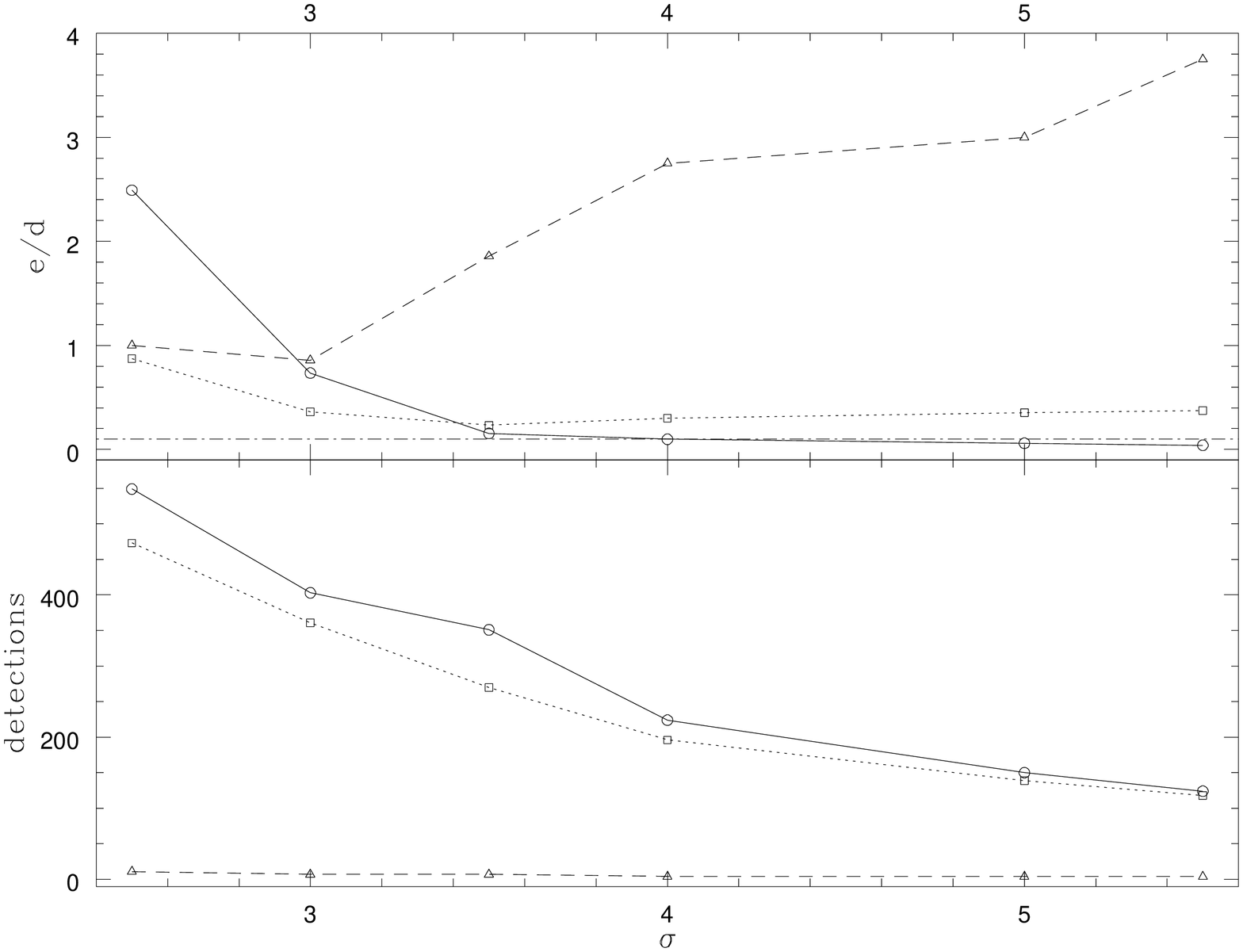}
\caption{Comparison of the performances of three
different filters over the 30 GHz TOD. Open circles and solid
line refer to optimal adaptive filter results. Boxes and
dashed line correspond to Mexican Hat Wavelet results. 
Triangles and dashed line correspond to Gaussian filter
results. Both the MHW and Gaussian filter have a width 
of $33\arcmin$. The lower panel shows the number of detections 
above the different $\sigma$ thresholds. The upper panel
shows the ratio between spurious and true detections
for the same   $\sigma$ thresholds.}
\label{fig5}
\end{figure*}

In the case we are considering
the bias is never greater than $16\%$.
 At intermediate thresholds, the (positive) effects of detection bias 
and the non-ideality of the data (negative) 
tend to cancel
and, {\it on average}, the amplitude estimates are unbiased.
To compare with other 'classical' filters we
repeated the process using a Gaussian filter
and  a Mexican Hat Wavelet (MHW, hereafter), both of them
with a  width equal to the beam width
of the antenna ($33\arcmin$).
The normalization of both filters was chosen
so that the coefficient at the position
of the source is equal to the amplitude of the source 
(that is, the filtering process does not
change the amplitude of the sources). The adaptive filter
automatically satisfies this condition (condition
2 for an optimal adaptive filter). 
While the MHW and the optimal adaptive filters are both
band-pass filters, the Gaussian is a low-pass filter, so
the comparison with the Gaussian is a bit unfair: the Gaussian is expected 
to perform significantly worse than the other two filters.
The number of detections above several thresholds
 for the two 'classical' filters together with 
the optimal adaptive filter are shown in table~\ref{tb1}. 
The number of detections is similar for the MHW and
the optimal adaptive filter, yet are slightly higher 
for the optimal adaptive filter.  The lowest number of detections
corresponds to the Gaussian filter. 
In the lower panel of figure~\ref{fig5} the number of
detections with the three different filters is compared. Detections
with the optimal adaptive filter are shown with open circles and
solid line. The open boxes and dashed line corresponds to MHW
detections and the triangles and dashed line corresponds to
Gaussian filter detections. The ratio between spurious and
true detections for the three filters is shown in the
top panel of figure~\ref{fig5}. Except for the $2.5\sigma$
level, the Gaussian filter produces the worst ratio. The 
optimal adaptive filter gives spurious to detected ($e/d$) ratios that quickly
decline with increasing $\sigma$ thresholds. The
$e/d$ ratio for the MHW remains almost constant with
$\sigma$ in the considered cases and clearly exceeds the
ratio obtained with the optimal adaptive filter. 
The m.r.a.e. and the bias
in the determination of the amplitude are 
huge in the case of the Gaussian filter.
Both have very similar values. That means
that the main source of error is systematic (the
filter is biased).
Considering the m.r.a.e,
the MHW seems to give amplitude estimates a few
percent better than the optimal adaptive filter.
Flux limits are similar in the MHW and
optimal adaptive filter cases. The Gaussian filter leads to
higher inferior flux and completeness limits.

Figure~\ref{fig2} provides a useful insight into
what is happening with the different filters.
The Gaussian filter smoothes the image, removing
very efficiently the small scale noise but
allowing the large structures (the Galaxy and
others extended fluctuations) to remain
in the image. The naive $\sigma$ thresholding
counts these bright, large structures as
sources, leading to a big relative number of spurious 
detections. Besides, the sources that by chance lie
on 'valleys' of the background can not be detected. On
the other hand, sources that lie on areas of positive
background are enhanced and can be more easily detected.
This can explain the large
and positive bias in the 
detections with the Gaussian filter.

Both MHW and optimal adaptive filter are
better prepared to deal with this problem than
the Gaussian window. Their profiles in
Fourier space drop to zero at low
frequencies and thus they are efficient
in removing large scale structures. Images
filtered with the MHW and the optimal
adaptive filter
in figure~\ref{fig2} are similar. A
visual inspection reveals that MHW smoothes
better the high-frequency fluctuations.
The optimal adaptive filter is more efficient
in removing medium and large structures.
The fluctuations around pixel 150 (corresponding
to one of the two observations of the Galaxy in
the ring) are better removed with the optimal
adaptive filter than with the MHW.
This is more apparent in figure~\ref{fig1} where
the profiles of the different filters in Fourier
space clearly indicate the faster drop of
the adaptive filter at low frequencies (large
scales) and its slower drop at high frequencies
(small scales).
The MHW can have the same problem as the Gaussian filter. 
However, the probability of this problem
should be
smaller due to its better efficiency in removing the Galaxy
and other large-scale structures. This effect explains
the higher $e/d$ ratio and the trend to positive bias.

We conclude that the optimal adaptive filter detects point sources better
than the MHW and the Gaussian window. The number of detections is
comparable to the number of detections with the MHW and clearly
higher than with the Gaussian window. The relative number
of spurious detections with the optimal adaptive filter 
is lower, except for very low detection thresholds,
than with the MHW and the Gaussian window.
Over $4\sigma$ the contamination of spurious detections 
is lower
than  $10\%$. At this level ($4\sigma$) the number of 
expected sources in all the sky is of a few hundreds (224 in 
our simulation) above fluxes of around 0.9 Jy.

\begin{figure*}
\includegraphics[angle=270, width=17cm]{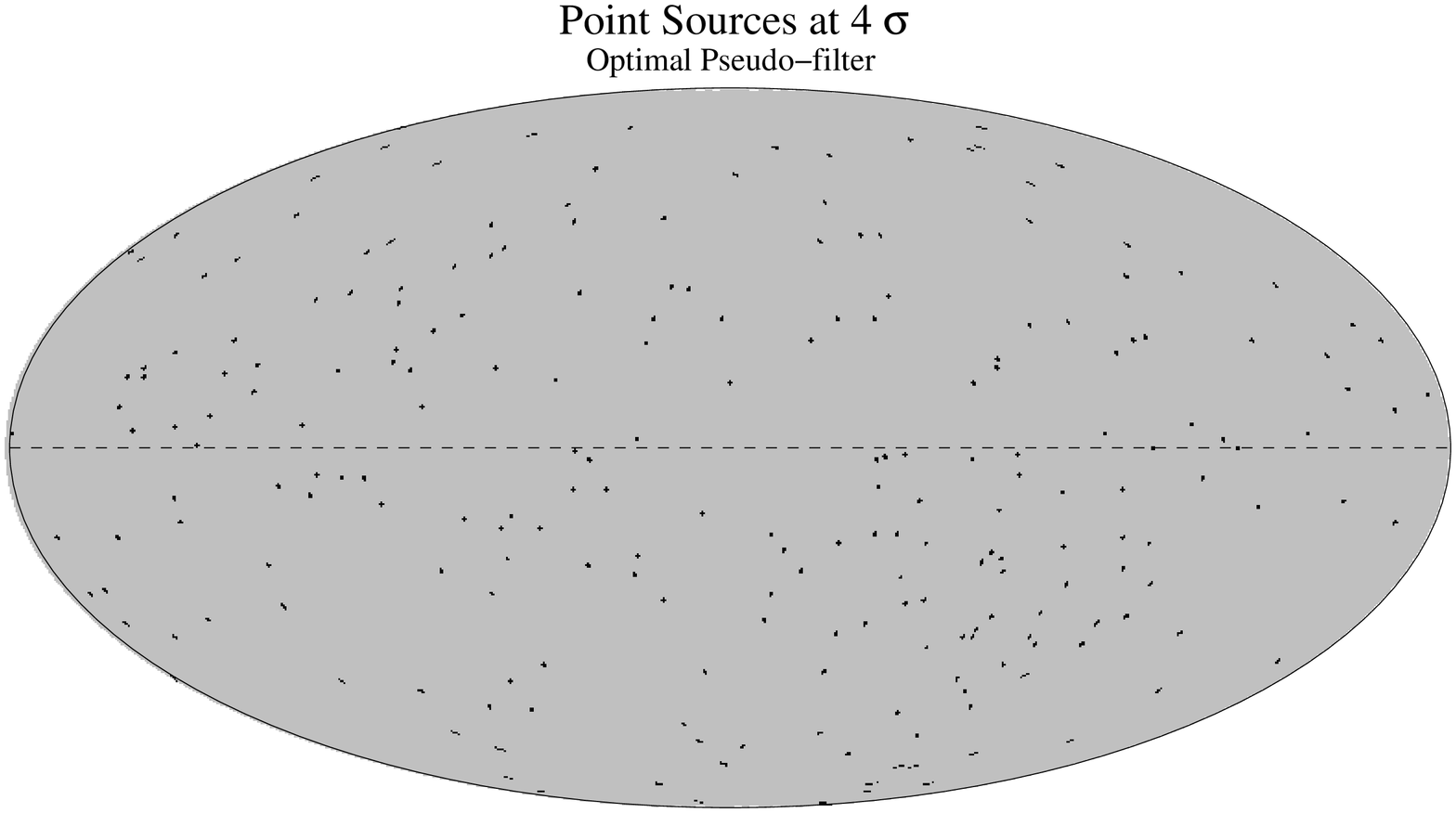}
\caption{Positions of the detected sources with
the optimal adaptive filter at a $4\sigma$ detection level. 
The sky is projected in Galactic coordinates. The dashed line represents 
the Galactic Equator. The optimal adaptive filter is able to detect
sources even in the highly contaminated region around the Galactic Plane.}
\label{fig6}
\end{figure*}

Some tests can be performed in order to discover if the number
of spurious sources, the most unpleasant effect of the filtering
and detection process, can be reduced.
Where do these spurious detections come from? One possibility is
that the peak finding algorithm is detecting the Galaxy or other
large-scale features. Such structure will appear in several
adjacent rings, as sources do, and therefore a
possibility of confusion exists. To test this potential source of 
contamination, we repeated the analysis excluding a band
of $5\deg$ centered on the Galactic plane (corresponding to
a $4.36\%$ of the sky area). The top panels of figures~\ref{fig3}
and~\ref{fig4} show the number of detections and the ratio of spurious/true
detections, respectively, for the optimal adaptive filter
in the cases where the rings have been averaged (as explained
before, filled circles) and where they have not (open circles). 
In the first section of table~\ref{tb2} are the results for the case
of averaged rings (the tabulated quantities are the same than
in table~\ref{tb1}) are shown.
The decrease in the
number of detections corresponds to the one expected for a uniform
distribution of sources in the sky (around  $5\%$). 
This indicates that the density of detections around the Galactic
plane is not substantially different from the density of detections
in other regions, less 'contaminated', of the sky. 
This can be seen in figure~\ref{fig6}, where the  $4\sigma$
detections
 have been represented in Galactic coordinates (the Galactic Plane
being represented by a dashed line).

The ratio
between spurious and true detections remains almost untouched   
for low $\sigma$. For higher $\sigma$ levels, the proportion 
of spurious detections that are due to the Galactic plane increases
dramatically (comparing  tables~\ref{tb1} and~\ref{tb2} we see that this
proportion increases from a $7\%$ at $2.5\sigma$ to a $89\%$ at  
$5\sigma$). This indicates that
the contamination by the Galaxy dominates at high signal to noise 
ratios, whereas low-intensity contamination is dominated by
noise fluctuations. In all cases, the number of peaks that
correspond to the Galaxy is much smaller than the number
of detected sources, thus implying that the optimal adaptive filter
deals efficiently with large scale backgrounds.

 \begin{table*}
 \centering
\begin{minipage}{122mm}
\caption{\label{tb2} Different tests for the detections at 30 GHz.
Columns have the same meaning than in table~\ref{tb1}.
Two different cases are tabulated:  detections   
with the optimal adaptive filter when a $5\deg$ band around 
the Galactic plane is excluded from the analysis  and detections with a $33\arcmin$
MHW  when a $5\deg$ band around 
the Galactic plane is excluded from the analysis. }

 \begin{tabular}{ c c c c c c c c }
\hline
  $\sigma$ & 
  detected & 
  spurious & 
  mean offset   & 
  m.r.a.e.  &
  $< bias >$ & 
  min. flux  & 
  compl. flux \\
	&
  sources &
  sources &
($\arcmin$) &
($\%$) &
($\%$) &
 (Jy) &
 (Jy) \\

\hline
\\
 \multicolumn{7}{c}{Optimal Adaptive filter, excluding the Galaxy} \\ 
\\
2.5  &    520  &    1272  &   14.06 & 23.79 &	 2.67  &  0.542 &   13.722 \\ 
3.0  &    379  &    257   &   13.51 & 23.69 &	 0.58  &  0.644 &   13.722 \\ 
3.5  &    280  &    34    &   12.67 & 20.87 &  	-6.04  &  0.760 &   13.722 \\ 
4.0  &    212  &    8     &   12.45 & 20.05 &  	-8.97 &   0.885 &   13.722 \\ 
5.0  &    142  &    1     &   12.16 & 18.73 &  -16.25  &  1.066 &   13.722 \\ 
5.5  &    118  &    1     &   11.94 & 18.97 &  -17.22 &  1.197  &   13.722 \\ 
\\
 \multicolumn{7}{c}{Mexican Hat Wavelet, excluding the Galaxy} \\ 
\\
2.5  &   454 & 347 & 12.76 & 20.21 &	2.79 & 0.531 & 13.72 \\ 
3.0  &   342 &  87 & 12.54 & 19.04 &	0.42 & 0.693 & 13.72 \\
3.5  &   256 &  25 & 12.49 & 18.28 &   -2.37 & 0.806 & 13.72 \\
4.0  &   187 &  14 & 12.15 & 15.71 &   -7.65 & 0.844 & 13.72 \\
5.0  &   129 &   7 & 12.18 & 15.07 &  -10.64 & 0.957 & 13.72 \\
5.5  &   108 &   7 & 12.01 & 14.60 &  -12.37 & 1.099 & 13.72 \\

\hline

 \end{tabular}
\end{minipage}
 \end{table*}

We stated before that 
the higher number of spurious detections of MHW
could be due to its non-optimal performance at
large scales. 
To further test this hypothesis we repeated the
test for the MHW, now considering only the peaks found outside
a $5\degr$ band centered in the Galactic plane, as we did with 
the optimal adaptive filter before. The results are shown in
the last section of table~\ref{tb2}.  Most of
the spurious sources (specially at high $\sigma$) lied 
near the Galactic plane, as we expected. And yet the remaining
number of spurious sources is still greater than in the
equivalent optimal adaptive filter case. The number of
detections and the flux limits remain similar to
the optimal adaptive filter case. The m.r.a.e. is also
similar in the two cases. 
Finally, the mean bias
in the determination of the amplitude is negative
for high $\sigma$ thresholds, as happens with the optimal
adaptive filter.  
Note that the MHW used in this work has the same scale
as the antenna. In fact, that scale is not the
optimal for detection (Vielva et al. 2000a). The optimal 
scale of the MHW for a particular case has to be
determined from the power spectrum of the data. This
in a certain way mimics the determination of the
optimal scale $R_{0}$ {\it that is automatically 
included} in the optimal adaptive filter via eqs. (\ref{eqfilter}) 
and (\ref{integrals}). The MHW at its optimal 
scales resembles the shape of the optimal adaptive filter
in Fourier space (in the sense that its maximum is located
near the maximum of the optimal adaptive filter) and thus
the effectiveness of both filters should be similar. 

    In order to further decrease the number of spurious detections,
we could take advantage of the fact that, in many
realistic cases, due to the sky coverage of the experiment and its 
scanning strategy, many positions of the sky
can be measured more than once (that is, at different epochs). 
For example, a source of $0\deg$ latitude will be detected
once when the center of the ring is located on longitude
$\phi=\phi_{source}-R$, being $R$ the radius of the ring,
and once again several months later, when the center of 
the ring is located on longitude
$\phi=\phi_{source}+R$.
When this occurs, it would be possible to almost duplicate
the amount of information in some areas of the sky and
therefore improve both the sensitivity and the reliability
of the detection. However, such a refined detection
can not be done when the
data do not cover the whole sky, and therefore is useless
for the construction of an early catalogue of sources
during the mission's flight.

\section{Conclusions}\label{conclusions}

We have simulated and analysed a sequence of time ordered data
such as the one that the future Planck 30 GHz LFI28 channel
will produce after the first 6 months of flight.
The data include all the main foregrounds as well as CMB
fluctuations, point sources and instrumental noise.
The resulting TOD has been ring-averaged in order to 
remove pixel-scale noise and then filtered with an
optimal adaptive filter that includes the spectral
properties of the data in order to maximize the 
detection of sources of a particular shape (Gaussian)
and scale (the scale of the antenna). The optimal
adaptive filter was designed to produce an unbiased,
efficient estimator of the amplitude of the sources 
at their position and to give a maximum of detections
at the characteristic scale of the sources. The detection
of the sources was performed by thresholding the
filtered TOD and looking for connected sets of peaks
belonging to adjacent rings. 
At a $4\sigma$ detection level 224 sources over a
flux of 0.88 Jy are detected with a mean relative error
(in absolute value) of $20.98\%$ and a systematic bias
of $-7.69\%$. The position of the sources in the sky 
is determined with errors inferior to the size of
the antenna. The catalogue of detected sources 
is complete at fluxes $\geq$ 4.337 Jy. 
The number of spurious detections is 22.

The performance of the optimal adaptive filter has been
compared with the performances under the same conditions 
of a Gaussian window and
a Mexican Hat Wavelet (MHW) of width equal to the beam width.
The number of sources detected with the optimal adaptive filter
is comparable to the number of sources detected with the MHW and
much higher than the number of sources detected with the Gaussian. 
However, the optimal adaptive filter finds a significantly lower
number of spurious sources than the other two filters. This is
due to the fact that the optimal adaptive filter removes better 
the large scale fluctuations (e.g. the Galaxy) than 
the other two filters. 
To test this hypothesis a $5\deg$ band around the
Galactic Plane was removed and the analysis was repeated with
the optimal adaptive filter and the MHW. This test 
shows that most of the spurious 
detections that were found with the MHW were located on the Galactic
Plane, whereas spurious detections that were
found with the optimal adaptive filter are uniformly distributed
in the sky.

The mean absolute error in the determination
of amplitudes is slightly higher in the case of the optimal 
adaptive filter than in the MHW case. The estimation
of the amplitudes with the Gaussian window suffer from very
large errors. In most cases, the Gaussian detects only the
Galaxy.

The number of spurious detections can be reduced by means
of further analysis after the filtering. 
For example, when the sky coverage of the scan is big enough 
some areas of the sky can be observed twice or more times,
increasing the signal to noise ratio of the sources that
lie in such areas.
This point is out 
of the scope of the present work, in which we only present
the filter as a first step in the data reduction. 

In conclusion, the optimal adaptive filter is an
efficient, unbiased and reliable tool for the detection
and extraction of punctual sources from TOD. A few hundred
of sources over 1 Jy will be detected in the 30 GHz channels 
of the future Planck mission with 
 $\leq 10 \%$ of spurious detections. 
The catalogue will be complete at fluxes $\geq 4$ Jy.
One possible application of this technique would be the 
elaboration of early catalogues of sources.
A greater number
of detections is expected at higher frequencies.
The simulation and analysis of such frequencies will be
performed in a future work.

\section*{Acknowledgments}

The authors strongly appreciate all the comments, suggestions and careful reading done by Matthias Bartelmann and Fabio Pasian.

The TOD used in this work for point source detection was 
generated by the Planck
pipeline simulator of the DPC Level-S.  The sky simulations, 
pointings and other
data is available for the Planck consortia at
http://www.mpa-garching.mpg.de/SimData.  We specially thank Matthias Bartelmann
and Klaus Dolag for their help in the generation of the simulations and their
support during this work.

The work presented here has used extensively the software package 
HEALPix
(Hierarchical, Equal Area and iso-Latitude Pixelisation of the sphere);
 we thank
the developing group:  Krzysztof M. Gorski, E.F. Hivon, Benjamin D. Wandelt, Antony J. Banday, F.K. Hansen and Matthias Bartelmann.  The HEALPix package is
available at this address:   http://www.eso.org/science/healpix

We thank Laura Cay\'on and Patricio Vielva for useful suggestions and comments.
DH acknowledges support from a Spanish MEC FPU fellowship. We thank FEDER Project 1FD97-1769-c04-01, Spanish DGESIC Project PB98-0531-c02-01 and INTAS Project INTAS-OPEN-97-1192 for partial financial support.

\end{document}